\documentclass[12pt]{article}
\usepackage{graphics,graphicx}
\usepackage{amssymb,epsfig,amsmath,euscript,array}
\usepackage[nosort]{cite}
\usepackage{axodraw4j}
\usepackage{pstricks}
\usepackage{color}

\makeatletter
\@addtoreset{equation}{section}
\makeatother

\newcounter{multieqs}

\newcommand{\be}{\begin{equation}}
\newcommand{\ee}{\end{equation}}

\newcommand{\bm}[1]{\mbox{\boldmath $#1$}}

\newcommand{\kslash}{k \!\!\! / }

\newcommand{\lslash}{l \!\! / }
\newcommand{\Pslash}{P \!\!\!\! / }

\newcommand{\islash}{i \!\!\! / }
\newcommand{\jslash}{j \!\!\! / }
\newcommand{\aslash}{a \!\!\! / }
\newcommand{\bslash}{{b \hspace{-6pt} \slash} }

\newcommand{\onslash}{1 \!\!\! / }
\newcommand{\twslash}{2 \!\!\!/ }
\newcommand{\thslash}{3 \!\!\!/ }
\newcommand{\foslash}{4 \!\!\! / }
\newcommand{\fislash}{5 \!\!\! / }

\newcommand{\mslash}{m \!\!\! / }

\def\bd{\begin{document}}
\def\ed{\end{document}}
\def\nn{\nonumber}
\def\bea{\begin{eqnarray}}
\def\eea{\end{eqnarray}}

\def\ab{(ijab)}
\def\ba{(ijba)}
\def\ijab{{\tr}_{+}(\islash\, \jslash\, \aslash \, \bslash)}
\def\ijba{{\tr}_{+}(\islash\, \jslash\, \bslash \, \aslash)}
\def\ijaP{{\tr}_{+}(\islash\, \jslash\, \aslash \, \Pslash)}
\def\ijPLa{{\tr}_{+}(\islash\, \jslash\, \Pslash_L \, \aslash)}
\def\ijaPL{{\tr}_{+}(\islash\, \jslash\, \aslash \, \Pslash_L)}
\def\ijPLza{{\tr}_{+}(\islash\, \jslash\, \Pslash_{L;z} \, \aslash)}
\def\ijaPLz{{\tr}_{+}(\islash\, \jslash\, \aslash \, \Pslash_{L;z})}
\def\ijPa{{\tr}_{+}(\islash\, \jslash\, \Pslash \, \aslash)}
\def\iaPb{{\tr}_{+}(\islash\, \aslash\, \Pslash \, \bslash)}
\def\ibPa{{\tr}_{+}(\islash\, \bslash\, \Pslash \, \aslash)}
\def\ijPmu{{\tr}_{+}(\islash\, \jslash\, \Pslash \, \mu)}
\def\ibmuP{{\tr}_{+}(\islash\, \bslash\, \mu \, \Pslash)}
\def\ibmua{{\tr}_{+}(\islash\, \bslash\, \mu \, \aslash)}
\def\iamub{{\tr}_{+}(\islash\, \aslash\, \mu \, \bslash)}
\def\jaPb{{\tr}_{+}(\jslash\, \aslash\, \Pslash \, \bslash)}
\def\ijmuP{{\tr}_{+}(\islash\, \jslash\, \mu \, \Pslash)}
\def\ijmum{{\tr}_{+}(\islash\, \jslash\, \mu \, \mslash)}
\def\ijmmu{{\tr}_{+}(\islash\, \jslash\, \mslash \, \mu)}
\def\ijmP{{\tr}_{+}(\islash\, \jslash\, \mslash \, \Pslash)}
\def\iabP{{\tr}_{+}(\islash\, \aslash\, \bslash \, \Pslash)}
\def\ijbP{{\tr}_{+}(\islash\, \jslash\, \bslash \, \Pslash)}
\def\jbPa{{\tr}_{+}(\jslash\, \bslash\, \Pslash \, \aslash)}
\def\ijPb{{\tr}_{+}(\islash\, \jslash\, \Pslash \, \bslash)}
\def\jbmua{{\tr}_{+}(\jslash\, \bslash\, \mu \, \aslash)}

\def\loablt{ {\tr}_{+}(\lslash_1\, \aslash \, \bslash\, \lslash_2)}

\def\ijlolt{{\tr}_{+}(\islash\, \jslash\, \lslash_1 \, \lslash_2)}
\def\ijltlo{{\tr}_{+}(\islash\, \jslash\, \lslash_2 \, \lslash_1)}
\def\ibloa{{\tr}_{+}(\islash\, \bslash\, \lslash_1 \, \aslash)}
\def\jaltb{{\tr}_{+}(\jslash\, \aslash\, \lslash_2 \, \bslash)}
\def\ialtb{{\tr}_{+}(\islash\, \aslash\, \lslash_2 \, \bslash)}
\def\bltloa{{\tr}_{+}(\bslash\, \lslash_2\, \lslash_1 \, \aslash)}
\def\jbloa{{\tr}_{+}(\jslash\, \bslash\, \lslash_1 \, \aslash)}
\def\ibPb{{\tr}_{+}(\islash\, \bslash\, \Pslash \, \bslash)}
\def\ijltb{{\tr}_{+}(\islash\, \jslash\, \lslash_2 \, \bslash)}

\def\ijloa{{\tr}_{+}(\islash\, \jslash\,  \lslash_1 \, \aslash)}
\def\ijblt{{\tr}_{+}(\islash\, \jslash\,  \bslash \, \lslash_2)}

\def\jakb{{\tr}_{+}(\jslash\, \aslash\, \kslash \, \bslash)}
\def\iakb{{\tr}_{+}(\islash\, \aslash\, \kslash \, \bslash)}

\def\tofo{{\tr}_{+}(\onslash\, \thslash\, \twslash \, \foslash)}
\def\foto{{\tr}_{+}(\onslash\, \thslash\, \foslash \, \twslash)}
\def\tofi{{\tr}_{+}(\onslash\, \thslash\, \twslash \, \fislash)}
\def\fito{{\tr}_{+}(\onslash\, \thslash\, \fislash \, \twslash)}

\def\lrangle#1#2{\langle #1\,#2\rangle}

\def\Li{{$\rm Li}_2$}
\def\eps{\epsilon}
\def\epsuv{{\epsilon_{\rm \mbox{\tiny UV}}}}
\let\bm=\bibitem
\let\la=\label

\def\npb#1#2#3{Nucl. Phys. {\bf{B#1}} #3 (#2)}
\def\plb#1#2#3{Phys. Lett. {\bf{#1B}} #3 (#2)}
\def\prl#1#2#3{Phys. Rev. Lett. {\bf{#1}} #3 (#2)}
\def\prd#1#2#3{Phys. Rev. {D \bf{#1}} #3 (#2)}
\def\cmp#1#2#3{Comm. Math. Phys. {\bf{#1}} #3 (#2)}
\def\cqg#1#2#3{Class. Quantum Grav. {\bf{#1}} #3 (#2)}
\def\nppsa#1#2#3{Nucl. Phys. B (Proc. Suppl.) {\bf{#1A}}#3 (#2)}
\def\ap#1#2#3{Ann. of Phys. {\bf{#1}} #3 (#2)}
\def\ijmp#1#2#3{Int. J. Mod. Phys. {\bf{A#1}} #3 (#2)}
\def\rmp#1#2#3{Rev. Mod. Phys. {\bf{#1}} #3 (#2)}
\def\mpla#1#2#3{Mod. Phys. Lett. {\bf A#1} #3 (#2)}
\def\jhep#1#2#3{J. High Energy Phys. {\bf #1} #3 (#2)}
\def\atmp#1#2#3{Adv. Theor. Math. Phys. {\bf #1} #3 (#2)}
\newcommand{\EQ}[1]{\begin{equation} #1 \end{equation}}
\newcommand{\AL}[1]{\begin{subequations}\begin{align} #1 \end{align}\end{subequations}}
\newcommand{\SP}[1]{\begin{equation}\begin{split} #1 \end{split}\end{equation}}
\newcommand{\ALAT}[2]{\begin{subequations}\begin{alignat}{#1} #2 \end{alignat}
                        \end{subequations}}
\def\beqa{\begin{eqnarray}}
\def\eeqa{\end{eqnarray}}
\def\beq{\begin{equation}}
\def\eeq{\end{equation}}
\def\sst{\scriptscriptstyle}
\def\thetabar{\bar\theta}
\def\Tr{{\rm Tr}}
\def\one{\mbox{1 \kern-.59em {\rm l}}}
 \def\Nh{\hat{N}}

\newcommand{\half}{{\textstyle {1 \over 2}}}

\def\a{\alpha}      \def\da{{\dot\alpha}}
\def\b{\beta}       \def\db{{\dot\beta}}
\def\c{\gamma}  \def\G{\Gamma}  \def\cdt{\dot\gamma}
\def\d{\delta}  \def\D{\Delta}  \def\ddt{\dot\delta}
\def\e{\epsilon}        \def\vare{\varepsilon}
\def\f{\phi}    \def\F{\Phi}    \def\vvf{\f}
\def\h{\eta}
\def\k{\kappa}
\def\l{\lambda} \def\L{\Lambda}
\def\m{\mu} \def\n{\nu}
\def\o{\omega}
\def\p{\pi} \def\P{\Pi}
\def\r{\rho}
\def\s{\sigma}  \def\S{\Sigma}
\def\t{\tau}
\def\th{\theta} \def\Th{\Theta} \def\vth{\vartheta}
\def\X{\Xeta}
\def\z{\zeta}
\def\de{\partial}

\def\cA{{\cal A}} \def\cB{{\cal B}} \def\cC{{\cal C}}
\def\cD{{\cal D}} \def\cE{{\cal E}} \def\cF{{\cal F}}
\def\cG{{\cal G}} \def\cH{{\cal H}} \def\cI{{\cal I}}
\def\cJ{{\cal J}} \def\cK{{\cal K}} \def\cL{{\cal L}}
\def\cM{{\cal M}} \def\cN{{\cal N}} \def\cO{{\cal O}}
\def\cP{{\cal P}} \def\cQ{{\cal Q}} \def\cR{{\cal R}}
\def\cS{{\cal S}} \def\cT{{\cal T}} \def\cU{{\cal U}}
\def\cV{{\cal V}} \def\cW{{\cal W}} \def\cX{{\cal X}}
\def\cY{{\cal Y}} \def\cZ{{\cal Z}}

\def\ua{\underline{\alpha}}
\def\ub{\underline{\phantom{\alpha}}\!\!\!\beta}
\def\uc{\underline{\phantom{\alpha}}\!\!\!\gamma}
\def\um{\underline{\mu}}
\def\ud{\underline\delta}
\def\ue{\underline\epsilon}
\def\una{\underline a}\def\unA{\underline A}
\def\unb{\underline b}\def\unB{\underline B}
\def\unc{\underline c}\def\unC{\underline C}
\def\und{\underline d}\def\unD{\underline D}
\def\une{\underline e}\def\unE{\underline E}
\def\unf{\underline{\phantom{e}}\!\!\!\! f}\def\unF{\underline F}
\def\unm{\underline m}\def\unM{\underline M}
\def\unn{\underline n}\def\unN{\underline N}
\def\unp{\underline{\phantom{a}}\!\!\! p}\def\unP{\underline P}
\def\unq{\underline{\phantom{a}}\!\!\! q}
\def\unQ{\underline{\phantom{A}}\!\!\!\! Q}
\def\unH{\underline{H}}

\def\As {{A \hspace{-6.4pt} \slash}\;}
\def\bs {{b \hspace{-6.4pt} \slash}\;}
\def\Ds {{D \hspace{-6.4pt} \slash}\;}
\def\ds {{\del \hspace{-6.4pt} \slash}\;}
\def\ss {{\s \hspace{-6.4pt} \slash}\;}
\def\ks {{ k \hspace{-6.4pt} \slash}\;}
\def\ps {{p \hspace{-6.4pt} \slash}\;}
\def\pas {{{p_1} \hspace{-6.4pt} \slash}\;}
\def\pbs {{{p_2} \hspace{-6.4pt} \slash}\;}
\def\Ps {{P \hspace{-6.4pt} \slash}\;}
\def\Qs {{Q \hspace{-6.4pt} \slash}\;}

\def\Fh{\hat{F}}
\def\Vh{\hat{V}}
\def\Xh{\hat{X}}
\def\ah{\hat{a}}
\def\xh{\hat{x}}
\def\yh{\hat{y}}
\def\ph{\hat{p}}
\def\xih{\hat{\xi}}
\def\psit{\tilde{\psi}}
\def\Psit{\tilde{\Psi}}
\def\tht{\tilde{\th}}
\def\lt{\tilde{\lambda}}
\def\hl{\hat{\lambda}}
\def\hlt{\hat{\tilde{\lambda}}}
\def\llt{\tilde{l}}
\def\At{\tilde{A}}
\def\Qt{\tilde{Q}}
\def\Rt{\tilde{R}}
\def\Nt{\tilde{N}}

\def\at{\tilde{a}}
\def\st{\tilde{s}}
\def\ft{\tilde{f}}
\def\pt{\tilde{p}}
\def\qt{\tilde{q}}
\def\vt{\tilde{v}}
\def\nt{\tilde{n}}

\def\delb{\bar{\partial}}
\def\bz{\bar{z}}
\def\bD{\bar{D}}
\def\bB{\bar{B}}

\def\bk{{\bf k}}
\def\bl{{\bf l}}
\def\bp{{\bf p}}
\def\bq{{\bf q}}
\def\br{{\bf r}}
\def\bx{{\bf x}}
\def\by{{\bf y}}
\def\bR{{\bf R}}
\def\bV{{\bf V}}

\def\d{\delta}\def\D{\Delta}\def\ddt{\dot\delta}
\def\pa{\partial} \def\del{\partial}
\def\xx{\times}
\def\uno{\mbox{1 \kern-.59em {\rm l}}}
\def\trp{^{\top}}
\def\inv{^{-1}}
\def\dag{{^{\dagger}}}
\def\pr{^{\prime}}
\def\lan{\langle}
\def\ran{\rangle}
\def\rar{\rightarrow}
\def\lar{\leftarrow}
\def\lrar{\leftrightarrow}
\newcommand{\0}{\,\!}      
\def\one{1\!\!1\,\,}
\def\im{\imath}
\def\jm{\jmath}
\newcommand{\tr}{\mbox{tr}}
\newcommand{\slsh}[1]{/ \!\!\!\! #1}
\def\vac{|0\rangle}
\def\lvac{\langle 0|}
\def\hlf{\frac{1}{2}}
\def\ove#1{\frac{1}{#1}}
\def\Box{\square}
\def\ZZ{\mathbb{Z}}
\def\CC#1{({\bf #1})}
\def\bcomment#1{}
\def\bfhat#1{{\bf \hat{#1}}}
\def\VEV#1{\left\langle #1\right\rangle}
\newcommand{\ex}[1]{{\rm e}^{#1}} \def\ii{{\rm i}}
\def\rr{{\rm r}} \def\rs{{\rm s}}\def\rv{{\rm v}}
\def\ri{{\rm i}}\def\rj{{\rm j}}
\newcommand{\lrbrk}[1]{\left(#1\right)}
\newcommand{\sfrac}[2]{{\textstyle\frac{#1}{#2}}}

\def\Li{{\rm Li}_2}

\font\mybb=msbm10 at 12pt
\def\bb#1{\hbox{\mybb#1}}

\font\myBB=msbm10 at 18pt
\def\BB#1{\hbox{\myBB#1}}

\long\def\symbolfootnote[#1]#2{\begingroup%
\def\thefootnote{\fnsymbol{footnote}}\footnote[#1]{#2}\endgroup}

\begin{document}

\begin{flushright}
QMUL-PH-12-11
\end{flushright}

\vspace{8pt}

\begin{center}

{\Large \bf All one-loop amplitudes  in  $\mathcal{N}=6$ superconformal }
\\
\vspace{0.4cm}
{\Large \bf Chern-Simons  theory  }

\vspace{16pt}

{\mbox {\bf  Andreas Brandhuber,  Gabriele Travaglini and   Congkao Wen}}%
\symbolfootnote[4]{
{\tt  \{ \tt \!\!\!a.brandhuber, g.travaglini, c.wen\}@qmul.ac.uk}
}

\begin{center}
{\small \em

Centre for Research in String Theory\\
School of Physics and Astronomy\\
Queen Mary University of London\\
Mile End Road, London E1 4NS, UK
}
\end{center}


\vspace{60pt} {\bf Abstract}
\end{center}

\noindent
We exploit a recently found connection between special triple-cut diagrams and tree-level recursive diagrams to derive a general formula capturing the multi-particle factorisation of arbitrary one-loop amplitudes in the ABJM theory.  This formula contains certain anomalous contributions which are reminiscent of the so-called non-factorising  contributions  appearing  in the factorisation of one-loop amplitudes in four-dimensional gauge theory. In the second part of the paper we derive a recursion relation for the supercoefficients of one-loop amplitudes in ABJM theory. By applying this recursion relation, any one-loop supercoefficient can be reduced to special triple-cut diagrams involving at least one four-point tree amplitude. In turn, this implies that any one-loop supercoefficient can be derived from tree-level recursive diagrams.

\setcounter{page}{0}
\thispagestyle{empty}
\newpage


\setcounter{tocdepth}{4}
\hrule height 0.75pt
\tableofcontents
\vspace{0.8cm}
\hrule height 0.75pt
\vspace{1cm}

\setcounter{tocdepth}{2}


\setcounter{footnote}{0}


\section{Introduction}

In this paper we aim to study aspects of  the scattering amplitudes in 
three-dimensional $\cN=6$ supersymmetric  Chern-Simons (matter) theory,
often referred to as ABJM theory \cite{abjm}.  It is a closely related cousin of $\cN=4$ super Yang-Mills (SYM) which  provides  us with a novel
example of the AdS/CFT duality, the conjecture being that  ABJM describes the low-energy physics of M2-branes near orbifold singularities.  It also  shares several properties with $\cN=4$ SYM although it is not maximally supersymmetric.
Common features include the existence of integrable systems for the anomalous dimensions of operators \cite{minahan1,minahan2},
classical integrability of the dual string theory   \cite{Arutyunov:2008if,Stefanski:2008ik,Sorokin:2010wn} and Yangian symmetry of scattering amplitudes \cite{Bargheer:2010hn,Huang:2010qy}.

Our focus will be on scattering amplitudes in ABJM theory, in particular tree-level and one-loop amplitudes and their unexpected relations. Superficially, amplitudes in ABJM and 
$\cN=4$ SYM  appear to be quite similar. They can be calculated with the same tools such as on-shell recursion relations \cite{Gang:2010gy}, (generalised) unitarity \cite{Chen:2011vv, Bargheer:2012cp, btw}, Grassmannians \cite{Lee:2010du} or twistor-string like formulae
\cite{Huang:2012vt}, 
and strikingly also exhibit dual conformal/Yangian symmetry \cite{Bargheer:2010hn,Huang:2010qy,Gang:2010gy, Lee:2010du}. The latter appears to be a consequence of the integrable, dual string model,  and its anomalous breaking \cite{Bargheer:2012cp, Bianchi:2012cq}, 
although less studied than in $\cN=4$ SYM, seems to follow a similar pattern.
But there are also marked differences -- the gluons are non-dynamical although they have interesting, residual physical effects  through their zero mode \cite{Bargheer:2010hn, Bargheer:2012cp}. Furthermore, due to the particular matter representations,  only amplitudes with even numbers of legs are non-zero and the infrared (IR) divergences are milder. In particular, tree- and one-loop amplitudes are IR finite and IR divergences appear first at two-loop order. Also the factorisation properties, which enter on-shell recursions and unitarity methods in an important way, display novel features which we will explore and exploit further below.

Compared to $\cN=4$ SYM,  relatively little is known about amplitudes in ABJM theory, although in the recent two years the situation has improved considerably. The four-point tree amplitude \cite{Agarwal:2008pu} seeds the BCFW recursion relation \cite{Gang:2010gy}, which allows in principle for the calculation of all tree amplitudes.  The four-point amplitude  at one loop was found to be vanishing in  
\cite{Agarwal:2008pu,Chen:2011vv}, unlike higher-point amplitudes.%
\footnote{Similarities with $\cN=8$ SYM in three dimensions have emerged recently in \cite{lionel}, where it has been shown that all one-loop MHV amplitudes in this  theory are vanishing.}  
In particular,  the six-point amplitude at one loop was explicitly computed in \cite{Bargheer:2012cp, Bianchi:2012cq, yutin}, while in \cite{btw} we constructed one-loop amplitudes up to ten points using a particular correspondence between special   ``anomalous" triple-cut diagrams and tree-level recursive diagrams. More concretely, an anomalous triple cut has a four-point amplitude as one of the   tree amplitudes appearing in the cut, in which case the triple cut is in one-to-one correspondence with a BCFW recursive diagram  where  the two external legs of the four-point amplitude 
in the triple cut, say $i$ and $i+1$, are mapped to shifted legs $\hat{i}$ and $\widehat{i+1}$ in the corresponding recursive diagram.   
Denoting the recursive diagram evaluated on the two physically distinct pole solutions in the BCFW recursion relation by $Y^{(1)}$ and $Y^{(2)}$,  the result of  \cite{btw} is  -- schematically --  
that  the triple-cut diagram is proportional to $Y^{(1)} -  Y^{(2)}$ multiplied by a combination of sign functions (while the tree-level recursive diagram is simply $Y^{(1)}+ Y^{(2)}$). 

This  correspondence is a close relative of the RSV formula \cite{dissolving}, which  expresses tree amplitudes in $\cN=4$ SYM as sums of two-mass hard coefficients, and points at a deep relations  between the S-matrix of ABJM at tree and one-loop level,%
\footnote{These similarities were firstly noticed in \cite{Bargheer:2012cp, Bianchi:2012cq}  and can be understood  in the six-point case from 
the anomalous  violation of Yangian invariance. }
which we will explore in great detail in this paper. 
 At two loops not much is known at present, the  only data point being the four-point amplitude  \cite{Chen:2011vv, Bianchi:2011dg}, 
whose expression surprisingly  matches that of the  one-loop amplitude in $\mathcal{N}=4$ SYM, even to  
all orders in the expansion in the dimensional regularisation parameter $\epsilon$  \cite{Bianchi:2011aa}.
Finally, Wilson loops and a  possible  duality  to amplitudes \cite{am,dks,bht} were  studied in 
\cite{Henn:2010ps,  Bianchi:2011rn, Wiegandt:2011uu, Bianchi:2011dg}.

The main focus of this paper is on the one-loop amplitudes in ABJM,  and in particular their intriguing connections to tree amplitudes observed in 
\cite{Bargheer:2012cp,Bianchi:2012cq,btw}. More specifically, we will concentrate on two distinct themes: the unexpected multi-particle factorisation properties at one-loop, and a new recursion relation for one-loop supercoefficients of the ABJM amplitudes.

An important and intriguing property of the ABJM amplitudes at one loop is their infrared finiteness  -- a fact that can be understood from the conjectured dual conformal invariance of the theory \cite{Bargheer:2012cp} or alternatively from the impossibility to cancel infrared divergences in physical quantities at one loop because of the absence of amplitudes with an odd number of legs \cite{btw}.
The finiteness of the one-loop S-matrix would naively lead to the conclusion that factorisation properties at this loop order should be trivial 
\cite{Bern:1995ix}. In particular there should be no ``non-factorising contributions" of the kind found in \cite{Bern:1995ix}  in four-dimensional gauge theory amplitudes, where the  peculiarity of these  contributions is that they contain kinematic invariants made of momenta  from both sides of the factorisation channel, as opposed to terms appearing in the naive factorisation.%
\footnote{The precise definition of trivial, or naive   factorisation is given in  \eqref{1f}.} 
 One then faces an immediate puzzle, discussed in Section 2, concerning the factorisation of the one-loop, six-point amplitude in three-particle channels. Indeed, the six-point amplitude at one-loop is proportional to a tree-level six-point amplitude \cite{Bargheer:2012cp,Bianchi:2012cq}, which has a non-trivial multi-particle factorisation in a three-particle channel. On the other hand, the vanishing of the one-loop four-point amplitude \cite{Agarwal:2008pu},  onto which the six-point amplitude naively factorises, would lead to the incorrect conclusion that the one-loop six-point amplitude is finite in the factorisation limit. 

We will be able to resolve this puzzle by resorting to the correspondence between three-particle cuts and BCFW tree-level  diagrams \cite{btw} described earlier. As explained in detail in Section 2, we will  show that in multi-particle limits some of the anomalous one-loop supercoefficients develop peculiar singularities which have a clear physical interpretation in terms of singularities of the associated BCFW  recursive diagrams. As a consequence, naive factorisation has to be augmented by non-factorising contributions; in  \eqref{one-final} we present the complete factorisation formula for all one-loop amplitudes in ABJM. 
Interestingly, our derivation of the non-factorising contributions from anomalous triple-cut diagrams (which, as recalled earlier are those  containing a four-point tree amplitude as one of the amplitudes participating in the cut) is tightly linked to the peculiar role of the  gluon zero-momentum  mode in the tree-level four-point amplitude, that has been pointed out by \cite{Bargheer:2012cp}.

In the second part of this paper we address the derivation of all one-loop supercoefficients of ABJM amplitudes, and hence of all superamplitudes at this loop order.  The key tool is a very simple recursion relation, derived in Section 3, that these coefficients  obey. Rather than deriving the recursion relations from factorisation, as done in \cite{Bern:2005hh} in four-dimensional gauge theory, we resort to a trick where, starting from a generic triple-cut diagram associated with a particular supercoefficient, we apply tree-level BCFW recursion directly to one of the tree amplitudes participating in the cut. In a generic theory, and specifically in four-dimensional Yang-Mills, this procedure  gives rise to certain diagrams which cannot be cast in the form of a recursion relation (see Figure 4(b) for an example).  It is a peculiarity of the ABJM amplitudes, which are non-vanishing only for an even number of legs, that  
such diagrams can always be avoided by appropriately choosing the legs to be shifted. As a consequence, a very simple recursion relation for supercoefficients can be derived which has the same form as the BCFW recursion for tree amplitudes. We give its final form  in \eqref{recsupercoeff}.  By repeatedly applying this recursion relation one can evaluate all one-loop supercoefficients in terms of anomalous triple-cut diagrams, and therefore in terms of tree-level recursive diagrams. The latter are then the  building blocks of all one-loop amplitudes.

The rest of the paper is structured as follows.  In Section 2 we derive the multi-particle factorisation properties of one-loop amplitudes exploiting the relation between BCFW tree-level diagrams and one-loop amplitudes in ABJM found recently by the authors in \cite{btw}. In Section 3 we show that in contrast to $\cN=4$ SYM all integral coefficients of one-loop amplitudes in ABJM obey on-shell recursion relations if appropriate legs are shifted. Remarkably this implies that the computation of all one-loop coefficients, and hence of all one-loop amplitudes, can essentially be reduced to tree-level on-shell recursions. In Section 4 we explain in some detail how these recursion relations  work in several examples and present a general, if somewhat formal, expression for the complete one-loop S-matrix. As a by-product our result establishes the Yangian invariance of all one-loop amplitudes except \cite{Bargheer:2012cp}
for very specific anomalies that have their origin in sign factors appearing in the one-loop amplitudes.

\section{Factorisation of one-loop amplitudes}

In this section we describe multi-particle factorisation of one-loop amplitudes in ABJM theory.
The case of one-loop six-point amplitudes  was first discussed in \cite{Bargheer:2012cp}, and 
in the following we will present a general factorisation formula valid for all one-loop 
amplitudes in ABJM. The key tool in our derivation is the correspondence found recently in \cite{btw}  between particular triple-cut diagrams, where at least one of the participating amplitudes is a four-point amplitude, and tree-level recursive diagrams, as we will describe shortly. 

We begin by briefly reviewing some basic facts about factorisation. The naive expectation is that amplitudes which do not have infrared divergences have trivial factorisation  \cite{Bern:1995ix}. Since one-loop amplitudes in ABJM theory are infrared finite, one would therefore assume  their factorisation to  be trivial. Schematically,%
\footnote{Note that in ABJM theory $n$ must be even and $j$ odd to have non-trivial multi-particle factorisation, since amplitudes with an odd number of legs vanish in this theory.}
\beq
\label{1f}
\cM^{(1)}_n \ \mathop{\sim}_{P^2_{1 j} \to 0 }\ \cM_{j+1}^{(1)} {1\over P^2_{1 j}} \cM_{n-j+1}^{(0)} \ + \  \cM_{j+1}^{(0)} {1\over P^2_{1 j}} \cM_{n-j+1}^{(1)}
\ , 
\eeq
where for the sake of definiteness we focus on the channel  $P_{1j} := p_1+ \cdots + p_j$, and 
$\cM^{(0)}$ ($\cM^{(1)}$) denotes a tree-level (one-loop) amplitude. 
By definition the right-hand side of this equation contains only the singular terms and all finite terms are dropped in the limit.
Similarly the expectation for tree amplitudes is to factorise as 
\beq
\label{2f}
\cM^{(0)}_n \ \mathop{\sim}_{P^2_{1j} \to 0 }\ \cM_{j+1}^{(0)} {1\over P^2_{1 j}} \cM_{n-j+1}^{(0)} 
\ .
\eeq
Two puzzles immediately arise with \eqref{1f} and \eqref{2f} when applied to ABJM amplitudes: 

{\bf 1.} The tree-level four-point amplitude $\cM_4^{(0)} (\bar{1}, \ldots , 4)$ has a pole $1/\lan12 \ran$ which arises in the forward-scattering limit $p_1 + p_2 \to 0$. This is  unaccounted for by \eqref{2f}, as there is no non-vanishing three-point amplitude in ABJM. 
However, we should emphasise that this is a rather special situation and is due to the gluon zero mode as reviewed later in this section. A generic collinear limit is of lower co-dimensionality and forces all four momenta to be collinear, in which case all four-point amplitudes vanish as expected.
 
{\bf 2.}
According to \eqref{1f}, the one-loop six-point amplitude  should factorise trivially in
a three-particle channel, {\it i.e.} 
\beq
\label{1loopsix}
\cM^{(1)}_6 \ \mathop{\sim}_{P^2_{13} \to 0 }\ \cM_{4}^{(1)} {1\over P_{13}^2} 
\cM_{4}^{(0)} \ + \  \cM_{4}^{(0)} {1\over P_{13}^2} \cM_{4}^{(1)} \ = \ 0
\ , 
\eeq
where the last equality follows since  $\cM^{(1)}_4 = 0$  \cite{Agarwal:2008pu}. 
However, it is known that%
\footnote{We follow the notation and conventions of Section 2 and Appendix A of \cite{btw} for the ABJM  superamplitudes and the three-dimensional spinor helicity formalism, respectively.}
\cite{yutin,Bargheer:2012cp,Bianchi:2012cq,btw}
\beq
\label{six}
\cM^{(1)}_6(\bar{1}, 2, \bar{3}, 4, \bar{5}, 6 ) \ = \  i \pi^3 \, \mathcal{S}(p) \,  \cM^{(0)}_6 (\bar{6}, 1, \bar{2}, 3, \bar{4}, 5 ) \, ,
\eeq
where the prefactor $\mathcal{S}(p)$ 
\beq
\mathcal{S}(p) \, =  \, {\rm sgn}( \lan 1 \, 2 \ran ) {\rm sgn}( \lan 3 \, 4 \ran ) {\rm sgn}( \lan 5 \, 6 \ran ) + 
{\rm sgn}( \lan 2 \, 3 \ran ) {\rm sgn}( \lan 4 \, 5 \ran ) {\rm sgn}( \lan 6 \, 1 \ran ) \, , 
\eeq
is a special combination of sign factors defined as
\beq
\label{segno}
{\rm sgn} \big( \lan k \, l \ran  \big)\ := \ - i  {\lan k \, l \ran \over \sqrt{ -(  \lan k \, l \ran^2+ i \varepsilon)}}
\ ,
\eeq
which is well defined for real and imaginary arguments.

Because the six-point tree amplitude on the right-hand side of \eqref{six} does have the nontrivial factorisation \eqref{2f} in the three-particle channel $P_{13}$, it follows that \eqref{1loopsix}, and hence \eqref{1f}, are also incomplete. Let us now proceed to identify the source of the problem and present its solution. 
%
%
%
\begin{figure}[h]
\centerline{\includegraphics[height=6cm]{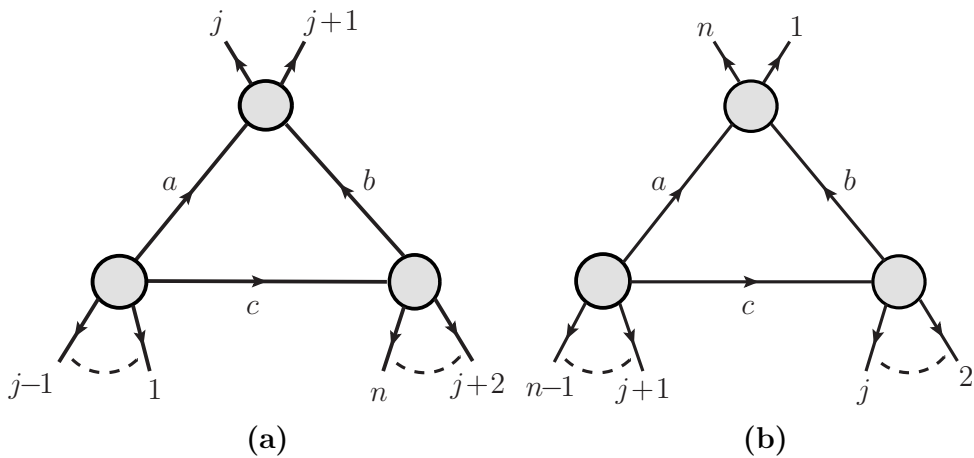} } 
\caption{\it
The particular three-particle cut diagrams giving rise to anomalous factorisation properties  of  one-loop amplitudes in the $P_{1j}^2 \to 0 $ limit.  }
 \label{fac-cut}
 \end{figure}
%
%

As before we focus on multi-particle factorisation of a one-loop amplitude in a kinematic channel containing an odd but otherwise arbitrary number of momenta $P_{1j}$. 
In the limit $P_{1j}^2 \to 0$ a generic triple-cut diagram contributing to the one-loop amplitude
remains either finite or contributes to the naive factorisation \eqref{1f}.
However, there are additional contributions from two special triple cuts, depicted in Figure \ref{fac-cut}, which also develop an unexpected simple pole of the form $1/P_{1j}^2$ in the factorisation limit. We will describe this now in detail focusing on the triple cut shown in Figure \ref{fac-cut}(a).  

\begin{figure}[h]
\centerline{\includegraphics[height=6cm]{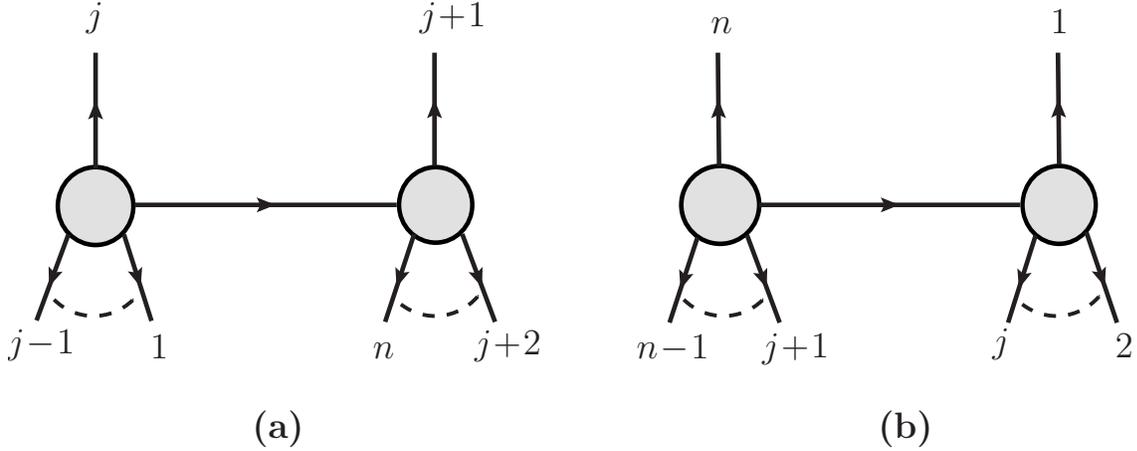} } 
\caption{\it
The two tree-level factorisation diagrams associated to the triple-cut diagrams in Figure \ref{fac-cut}.  }
 \label{fac-rec}
 \end{figure}

To begin our discussion, we recall the main result of \cite{btw}, namely the fact that triple-cut diagrams containing a four-point amplitude can be associated with (and calculated in terms of the residues of)  particular tree-level recursive diagrams. Specifically, the diagram 
in Figure  \ref{fac-cut}(a) can be associated with the tree-level recursive  diagram  represented  in Figure \ref{fac-rec}(a). 
The main idea can be conveyed schematically as follows. 

%
%
%
%
%

The supercoefficient associated with the triple cut of Figure \ref{fac-cut}(a) is equivalent to the recursive diagram shown in Figure \ref{fac-rec}(a). The crucial observation is that 
in the factorisation limit $P_{1j}^2 \to 0$ the usual shifts of the external momenta implied in the BCFW diagram are removed, since the internal propagator in Figure \ref{fac-rec}(a) goes on shell as $P_{1j}^2 \to 0$. Therefore, 
in this limit the supercoefficient  can be written schematically as  
\beq
\cC_{j j+1; n } \mathop{\sim}_{P_{1j}^2 \to 0 } \cM_L^{(0)} {1\over P_{1j }^2  }\cM_R^{(0)}
\ . 
\eeq
The triple-cut diagram in Figure \ref{fac-cut}(b) gives a similar contribution. The final result has the form 
\beq
\label{non-fact}
\mathcal{M}_L^{(0)}{\cF \over P_{1j}^2} \mathcal{M}_R^{(0)}
\, , 
\eeq
where the coefficient function $\cF$ will be determined below. 
This anomalous factorisation term has the structure of a product of two tree-level amplitudes multiplied by a propagator and the coefficient function $\cF$.
Importantly, this function depends on kinematic invariants with 
external momenta from both sides of the factorisation channel, 
and hence  is less universal than the naive factorisation contributions.
We note that \eqref{non-fact} has the same  form of the  so-called ``non-factorising terms" discussed in \cite{Bern:1995ix} in the context of four-dimensional gauge theories.

Let us now fill in the details omitted in the previous qualitative discussion, and derive the general formula capturing the factorisation of all one-loop amplitudes in ABJM.   In order to do so, we briefly review the results of  \cite{btw}, focusing again on the triple-cut diagram of Figure \ref{fac-cut}(a). 
It was found  in \cite{btw} that the  cut momenta  $(l_a)_{\a \b}  := \hat{\l}_{a; \a}\hat{\l}_{a; \b}$ and $(l_b)_{\a \b}  := \hat{\l}_{b; \a}\hat{\l}_{b; \b}$ 
of the  triple-cut diagram of Figure  \ref{fac-cut}(a) can be expressed  in terms of the spinors 
$\hat\l_a$ and  $\hat\l_b$ defined as
\beq
\label{shiftl}
\begin{pmatrix} 
\hat\l_a \\ \hat\l_b 
\end{pmatrix} \ = \ 
R(z) \, \begin{pmatrix} 
\l_j \\ \l_{j+1}
\end{pmatrix}
\ , 
\eeq
where $R(z)$ is a rotation matrix, parameterised as 
\beq
R(z) \ = \ 
\begin{pmatrix} 
{1\over 2} (z + z^{-1})  & - {1\over 2i }  (z - z^{-1}) 
\\ \cr
{1\over 2i }  (z - z^{-1})  & 
{1\over 2}  (z + z^{-1}) 
\end{pmatrix} 
\ . 
\eeq
The shift parameter $z$, which is the analogue of the deformation parameter $z$ of four-dimensional BCFW recursion relations, is fixed by solving the remaining on-shell condition 
\beq
\label{remai}
l_c^2 =(l_a + K_1)^2 = 0
\ , 
\eeq
with $K_1 = P_{1\, j-1}$. 
It was shown in \cite{Gang:2010gy} and also in Section  3 of \cite{btw} that this condition can be put in the form 
\beq
\label{biquadratica}
a z^{-2} + b + c z^2 = 0 
\ , 
\eeq
with
\beq
\label{coeff}
a \ = \ 2 (\tilde{q} \cdot K_1) \ , \qquad b \ = \ - K_1 \cdot K_2 \ , \qquad c \ = \ 2 (q \cdot K_1) \ , 
\eeq
where
\beq
q^{\a \b} \ := \ {1\over 4} (\l_j + i \l_{j+1})^\a (\l_j + i \l_{j+1})^\b \ , 
\qquad 
\tilde{q}^{\a \b} \ := \ {1\over 4} (\l_j - i \l_{j+1})^\a (\l_j - i \l_{j+1})^\b \ , 
\eeq
and $K_2 := P_{j+2\, n}$. 
In this notation 
\beq
l_a \ := \ \hat{\l}_a \hat{\l}_a  \ = \ z^2 q \, +\,  z^{-2} \tilde{q} \, + \, {1\over 2} ( p_j + p_{j+1}) 
 \ , 
 \eeq
and the explicit solutions are
\beq
\label{explsol}
z_1^2 \ = \ {  K_1 \cdot K_2 + \sqrt{ K_1^2 K_2^2} \over 4  (q \cdot K_1) } \,  , \qquad 
z_2^2 \ = \ {  K_1 \cdot K_2 - \sqrt{ K_1^2 K_2^2} \over 4  (q \cdot K_1) }
\ .
\eeq
The supercoefficient in Figure \ref{fac-cut}(a)  is then given by \cite{btw}
\beqa
  \label{supxm}
  \cC_{j j+1; n}  & = &   
-\lan  j j+1 \ran \sqrt{K_1^2 K_2^2}\, 
\Big( Y_{j j+1 ; n}^{(1)} - Y_{j j+1; n}^{(2)} \Big)
 \ ,  
  \eeqa
where $Y_{j j+1; n}^{(a)}$, $a=1,2$ is the result of the recursive diagram in Figure \ref{fac-rec}(a) evaluated on the solution $z=z_a$, {\it i.e.}  \cite{Gang:2010gy}  
\beq
Y_{j j+1; n }^{(1)} \ =  \ 
 \int\!d^3 \eta_c \   {H(z_1, z_2) \over P_{1j}^2}  \Big[ \cM_R(\overline{j+2}  \ldots , n , - \bar{c} , \widehat{j+1}    )   \, \cM_L(\overline{1}, \ldots , j-1 , \bar{\hat{j}}, c    )  \Big]_{z = z_1} 
 \ , 
 \eeq
with $Y_{j j+1; n}^{(2)} \ = \ [ Y_{j j+1; n}^{(1)}]_{z_1 \leftrightarrow z_2}$, and the function $H$ is defined as 
\beq
H(z_1, z_2) := {z_1 ( z_2^2 - 1 ) \over z_1^2 - z_2^2}
\ . 
\eeq
Rewriting the off-shell momenta $K_1$ and $K_2$ in three-dimensional spinor notation as
\beq
\label{K1K2}
K_{1\, ab}\  := \ \xi_{(a} \mu_{b)}\, , \qquad  K_{2\, ab} \ :=\  \xi_{(a}^\prime \mu_{b)}^\prime\, , 
\eeq
with
 $ K_1^2 K_2^2 =(1/16) \lan \xi \mu \ran^2 \lan \xi^\prime \mu^\prime \ran^2$, 
we  can re-express \eqref{supxm} as follows: 
\beqa
  \label{supxm-bis}
  \cC_{j j+1; n}  & = &   
{\lan j j+1\ran\over 4}  \lan \xi \mu\ran \lan \xi^\prime \mu^\prime \ran \, \Big( Y_{j j+1; n}^{(1)} - Y_{j j+1; n}^{(2)} \Big) 
 \ . 
  \eeqa
We are interested in finding the behaviour of the triple-cut diagram 
of Figure \ref{fac-cut}(a) in the multi-particle factorisation limit 
$P_{1j}^2 \to 0$. In this limit, the coefficients $a$, $b$, $c$ introduced in \eqref{coeff} satisfy the relation $a + b + c \to 0$, from which one infers that 
\beq
z_2^2 \to 1 
\ . 
\eeq
Curiously, the specific limiting value of $z_1$ will be  immaterial in the following discussion. 

Next we take the factorisation limit $P_{1j}^2 \to 0$ on  \eqref{supxm-bis}. In this limit
\beq
{H(z_1, z_2)} \to 0\, , \qquad 
{H(z_2, z_1)} \to -1 \ , 
\eeq
hence only the second term in \eqref{supxm-bis} survives:
  \beq
  \label{coefflimit}
  \cC_{j j+1; n }  \, \mathop{\to}_{P_{1j}^2 \to 0 } \, 
{\lan j j+1\ran\over 4}  \lan \xi \mu\ran \lan \xi^\prime \mu^\prime \ran  \, 
 \int\!d^3 \eta_c \   {1\over P_{1j}^2}  \Big[ \cM_R^{(0)}(\overline{j+2}  \ldots , n, - \bar{c} , j+1    )   \, \cM_L^{(0)}(\overline{1}, \ldots , j-1 , \bar{j}, c    )  \Big]
 \ ,   
  \eeq
where we note that we were able to remove the BCFW shifts from legs $j$ and $j+1$, as these are turned off when $z\to 1$. 
  
Finally, the contribution to the factorisation of the amplitude is obtained by  multiplying \eqref{coefflimit}  
with the corresponding three-mass triangle function
\beqa
\label{3mts}
\cI^{\rm 3m}(K_1, K_2, K_3) &:= &  \int\!\!{d^3 l}  \, {1 \over( l^2 + i \varepsilon)  ( (l +K_1)^2 + i \varepsilon) ( (l +K_1 + K_2)^2 + i \varepsilon)  }
\nonumber  \\
&=& { -i \,  \pi^3 \over  \, \sqrt{- (K_1^2+ i \varepsilon)}   \sqrt{ -(K_2^2+ i \varepsilon)}  \sqrt{-( K_3^2 + i \varepsilon)} }
\  ,
\eeqa
with $K_3 = p_j + p_{j+1}$ in the  case at hand. Doing so we arrive at the  result 
\beqa
\label{afterlimit-final-a}
&&\cC_{j \, j+1; n } \, \cI (P_{j+2 \, n } , P_{1\,  j-1}, P_{j\,  j+1} )  \mathop{\to}_{P_{1j}^2 \to 0 }  
\nonumber \\
&&-i{  \pi^3\over 4}  { \lan j j+1\ran \over \sqrt{- (P_{j j+1}^2 + i \varepsilon)}}
{\lan \xi \mu \ran  \over  \sqrt{- (P_{1\,  j-1}^2 + i \varepsilon)}} { \lan \xi^\prime \mu^\prime \ran  \over   \sqrt{- (P_{j+2 \, n}^2 + i \varepsilon)}}
 \nonumber \\ \cr
&&\times \int\!d^3 \eta_c \   {1\over P_{1j}^2}  \Big[ \cM_R(\overline{j+2}  \ldots , n , - \bar{c} , j+1    )   \, \cM_L(\overline{1}, \ldots , j-1 , \bar{j}, c    )  \Big]
%
%
\ , 
\eeqa
where we set $( P_{1\,  j-1})_{\a \b} := \xi_{(\a} \mu_{\b)}$, and $(P_{j+2 \, n })_{\a \b} := \xi^\prime_{(\a} \mu^\prime_{\b)}$.

There is another contribution to add, namely that of Figure \ref{fac-cut}(b). This can be associated with the recursive diagram in Figure \ref{fac-rec}(b) and is given by
\beqa
\label{afterlimit-final-b}
 &&\cC_{n 1; j } \, \cI (P_{n  \, 1}, P_{j+1 \, n-1}, P_{2\,  j}) \mathop{\to}_{P_{1j}^2 \to 0 }  
 \nonumber \\
&&-i{  \pi^3\over 4}  { \lan n  1\ran \over \sqrt{- (P_{n 1}^2 + i \varepsilon)}}
{\lan \tilde\xi \tilde\mu \ran  \over  \sqrt{- (P_{2\,  j}^2 + i \varepsilon)}} { \lan \tilde{\xi}^\prime \tilde{\mu}^\prime \ran  \over   \sqrt{- (P_{j+1 \, n-1}^2 + i \varepsilon)}}
\nonumber  \\ \cr
 &&\times \int\!d^3 \eta_c \   {1\over P_{1j}^2}  \Big[ \cM_R(\overline{j+2}  \ldots , n, - \bar{c} , j+1    )   \, \cM_L(\overline{1}, \ldots , j-1 , \bar{j}, c    )  \Big] \, 
\eeqa
with $(P_{2\,  j})_{\a \b} := \tilde{\xi}_{(\a}  \tilde{\mu}_{\b)}$ and $ (P_{j+1 \, n-1})_{\a \b}    := \tilde{\xi}^\prime_{(\a} \tilde{\mu}^\prime_{\b)}.$
Hence the total anomalous factorisation term is obtained by summing \eqref{afterlimit-final-a} and \eqref{afterlimit-final-b}, and reads 
\beq
\label{afterlimit-final}
  \mathcal{F} \times \int\!d^3 \eta_c \   {1\over P_{1j}^2}  \Big[ \cM_R^{(0)}(\overline{j+2}  \ldots , n, - \bar{c} , j+1    )   \, \cM_L^{(0)}(\overline{1}, \ldots , j-1 , \bar{j}, c    )  \Big]
  \ , 
 \nonumber
\eeq
where
\beqa
\label{cal-F}
\mathcal{F} &= &-i{  \pi^3\over 4} \left[ 
 { \lan j j+1\ran \over \sqrt{- (P_{j j+1}^2 + i \varepsilon)}}
{\lan \xi \mu \ran  \over  \sqrt{- ( P_{1\,  j-1}^2+ i \varepsilon)}} { \lan \xi^\prime \mu^\prime \ran  \over   \sqrt{- (P_{j+2 \, n}^2 + i \varepsilon)}} 
\right.
\ + \ 
\nonumber \\ \cr
&&+  
\left.
 { \lan n  1 \ran \over \sqrt{- (P_{n 1}^2 + i \varepsilon)}}
 {\lan \tilde\xi \tilde\mu \ran  \over  \sqrt{- (P_{2\,  j}^2 + i \varepsilon)}} { \lan \tilde{\xi}^\prime \tilde{\mu}^\prime \ran  \over   \sqrt{- (P_{j+1 \, n-1}^2 + i \varepsilon)}}
\right] 
\ . 
\eeqa
In summary, the complete factorisation formula for one-loop amplitudes in ABJM theory is given by%
\footnote{In the following formula, integration over $ \eta_c$ is understood.}
\beqa
\label{one-final}
\cM^{(1)}_n & 
\stackrel{P^2_{1 j} \to 0 }{\longrightarrow}& \cM_{j+1}^{(1)} {1\over P^2_{1 j}} \cM_{n-j+1}^{(0)} \ + \  \cM_{j+1}^{(0)} {1\over P^2_{1 j}} \cM_{n-j+1}^{(1)}
\nonumber \\
&&+ \ 
\cM_{j+1}^{(0)}\,  {\mathcal{F}\over P^2_{1 j}} \, \cM_{n-j+1}^{(0)} 
\ , 
\eeqa
where $\cF$ is given by \eqref{cal-F}. The first line of \eqref{one-final} captures the naive factorisation, while the second line represents the non-factorising term. 

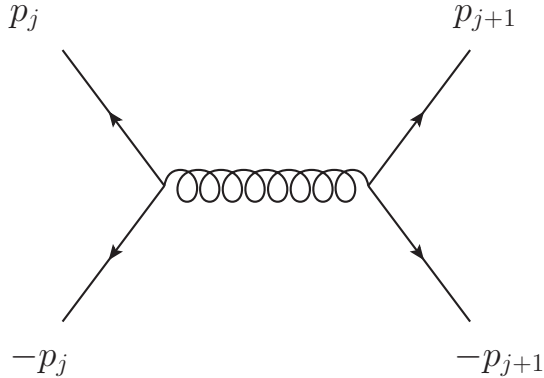
\begin{figure}[h]
\begin{center}
\scalebox{0.8}{
\fcolorbox{white}{white}{
  \begin{picture}(335,193) (164,-109)
    \SetWidth{1.0}
    \SetColor{Black}
    \Line[arrow,arrowpos=0.5,arrowlength=5,arrowwidth=2,arrowinset=0.2](272,-15)(224,49)
    \Line[arrow,arrowpos=0.5,arrowlength=5,arrowwidth=2,arrowinset=0.2](368,-15)(416,49)
    \Line[arrow,arrowpos=0.5,arrowlength=5,arrowwidth=2,arrowinset=0.2](272,-15)(224,-79)
    \Line[arrow,arrowpos=0.5,arrowlength=5,arrowwidth=2,arrowinset=0.2](368,-15)(416,-79)
    \Gluon(272,-15)(368,-15){7.5}{8}
    \Text(200,58)[lb]{\Large{\Black{$p_j$}}}
    \Text(200,-105)[lb]{\Large{\Black{$-p_j$}}}
    \Text(410,58)[lb]{\Large{\Black{$p_{j+1}$}}}
    \Text(410,-105)[lb]{\Large{\Black{$-p_{j+1}$}}}
  \end{picture}
}
}
\end{center}
\caption{\it
The dominant Feynman diagram contributing to the forward-scattering limit of the four point-amplitude $\cM_4^{(0)}(-p_j, p_j, p_{j+1}, - p_{j+1})$. The exchanged gluon has vanishing momentum.  }
 \label{FD}
 \end{figure}

Let us briefly pause here to discuss the subtle role played by the gluon zero-modes. 
It was suggested in \cite{Bargheer:2012cp} that 
the non-factorising terms are associated with the propagation of a gluon zero-momentum mode. This is precisely what emerges from the analysis presented in this section. 
Focusing again on the triple cut  in Figure \ref{fac-cut}(a), we observe that as $P_{1j}^2 \to 0$, the cut momenta $l_a$ and 
$l_b$ tend to the limiting values
\beq
l_a \rightarrow p_j \, , \qquad l_b \rightarrow p_{j+1}\, ,\quad  {\rm as} \ \ P_{1j}^2 \to 0
\ . 
\eeq
This is nothing but the well-known forward-scattering limit of the four-point amplitude $\cM^{(0)} (-l_a, - l_b, p_j, p_{j+1})$. In this limit the amplitude is singular \cite{Bargheer:2010hn},  with the singularity coming from the particular Feynman diagram drawn in Figure \ref{FD}, where a pair of (unscattered) particles exchanges a gluon with zero  three-momentum.


We conclude this section by applying our general result on  one-loop factorisation
to the simple example of the  one-loop six-point amplitude. 
More concretely, we consider the multi-particle factorisation  channel $P_{13}^2 \to 0$.  
As discussed earlier, there is no one-loop four-point amplitude in ABJM theory, hence the first line of \eqref{one-final} vanishes. 
In this case there are two different three-particle cut diagrams contributing to the non-factorising terms,
 where the external legs are grouped as $(61)$, $(23)$, $(45)$ and $(34)$, $(56)$, $(21)$.  The non-factorising term $\cF$ given in \eqref{cal-F} is easily found to be
\beq
 \cF_{P_{13}^2}  \ = \   \pi^3 \Big[ {\rm sgn} (\lan 61\ran)  {\rm sgn}(\lan23\ran){\rm sgn} (\lan 45\ran) \, + \,  {\rm sgn}(\lan 3 4 \ran) {\rm sgn}(\lan 56\ran){\rm sgn}( \lan 12\ran )\Big]
\ , 
\eeq
where we have  used the relations $\lan \xi \mu\ran = -2i \lan12\ran$, $\lan \xi^\prime \mu^\prime\ran = -2i \lan56\ran$, 
$\lan \tilde{\xi} \tilde\mu\ran = -2i \lan45\ran$, and
$\lan \tilde{\xi}^\prime \tilde{\mu}^\prime\ran = -2i \lan23\ran$, and ${\rm sgn} (\lan ij\ran)$ is defined in \eqref{segno}. 

As a final comment we would like to add that it would be tempting to construct a BCFW style recursion for one-loop amplitudes in ABJM theory based on our complete understanding of factorisation \eqref{one-final}. We have not attempted this here and leave this for future studies. Note that in the next section we will follow a slightly different route and introduce recursion relations for integral coefficients which allow us to construct at least in principle
the complete one-loop S-matrix of ABJM theory.

\section{Recursion relation for supercoefficients}
\label{BCFW}

The coefficients of  $L$-loop amplitudes are rational functions just as  tree-level amplitudes, and hence it is natural to consider  on-shell recursion relations for them. This idea was applied for the first time to one-loop amplitudes in four-dimensional gauge theory in \cite{Bern:2005hh}. 
However, one has to face two potential problems: firstly, individual integral coefficients may have spurious poles which have to cancel in the complete amplitude; 
and furthermore, it is not known a priori if the coefficients have the desired large-$z$ behaviour under BCFW shifts. 
However, in \cite{Bern:2005hh} a set of criteria has been derived 
under which recursion relations can be applied directly to coefficients. 

\begin{figure}[h]
\scalebox{1}{
\centerline{\includegraphics[height=4.5cm]{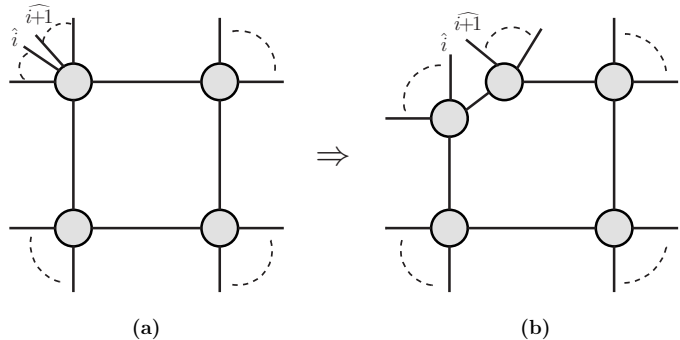} }
}
\caption{\it 
In (b) we show a possible diagram which may appear when applying the BCFW recursion relation to the top left tree amplitude in the quadruple cut shown in (a).  Such a situation cannot be avoided in general  in $\mathcal{N} = 4$ SYM, and  
the diagram in (b) cannot be described using  recursion relations for supercoefficients. 
$\hat{i}$ and $\widehat{i+1}$ denote the shifted legs.  }
 \label{fig1}
 \end{figure}

An elegant way to avoid all the above mentioned problems in one stroke is to use BCFW shifts of two legs that sit at the same corner of a cut loop diagram  \cite{Bern:2005hh}. 
This effectively relates the recursion relation for a coefficient to the recursion relation
for a tree amplitude (which we fully understand) which appears as one factor in the expression for the coefficient obtained from generalised unitarity.
Hence  the  knowledge of tree-level amplitudes allows us to determine the poles as well as the large-$z$ behaviour. However some more care is needed,  since the recursion relation will in general include diagrams such as that in Figure \ref{fig1}(b), where the propagator of the BCFW diagram is part of the (cut) loop diagram.
In this case $z$ dependence would enter the loop integration, lead to $z$-dependent spurious singularities, and destroy our attempt to construct an on-shell recursion relation for coefficients.
This is exactly the reason why there is no simple BCFW recursion relation for general supercoefficients in $\mathcal{N}=4$ SYM. 
On the other hand, if we are able to avoid channels of the type depicted in Figure \ref{fig1}(b), then the  recursion relation for coefficients follows immediately from that  for tree-level amplitudes. 

\begin{figure}[h]
\scalebox{1}{
\centerline{\includegraphics[height=4.5cm]{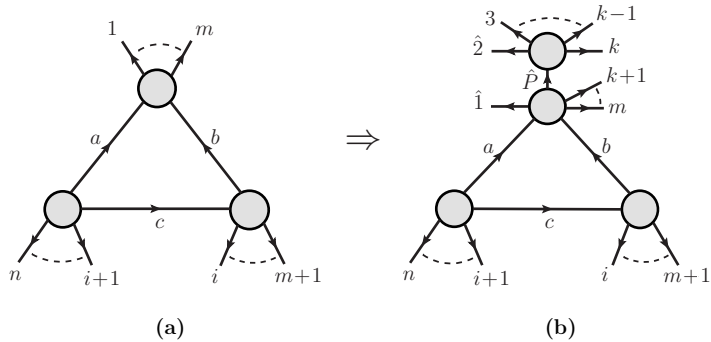} }
}
\caption{\it
In (b) we show a  recursive diagram for the  supercoefficient   $\mathcal{C}_{n; 1, \ldots, m; i }$  of a  one-loop amplitude in ABJM, calculated by the triple cut in (a).  }
 \label{fig2}
 \end{figure}

As we will now demonstrate ABJM theory does have such recursion relations for all one-loop supercoefficients.%
\footnote{The authors of  \cite{Bern:2005hh} were able to find valid  recursion relations for bubble and triangle coefficients in four-dimensional gauge theories.} 
The crucial property of ABJM that makes this possible is that all amplitudes with an odd number of particles vanish. This immediately implies that a recursive diagram such as that in Figure \ref{fig1}(b) can always be avoided by choosing appropriate locations for the shifted legs (labelled by  $\hat{i}$  and $\widehat{i+1}$ in Figure \ref{fig1}). 
Let us consider the concrete example in Figure \ref{fig2}, where the supercoefficient is given by the 
triple cut averaged over the two inequivalent solutions $l_{a,s}$, $s=1,2$ for the cut momentum $l_a$,  
\beqa \label{general}
\mathcal{C}_{n; 1, 2, \ldots, m; i} &=& {1\over 2} \sum_{s=1}^2 \int\!d^3 \eta_a d^3 \eta_b d^3 \eta_c  \,
\mathcal{M}^{(0)}(\bar{1}, \ldots, m, -\bar{b}_s, -a_s) 
\nonumber \\ \cr
&\times & \mathcal{M}^{(0)}(\overline{m\!+\!1}, \ldots, i, -\bar{c}_s, b_s)
 \mathcal{M}^{(0)}(\overline{i\!+\!1}, \ldots, n, \bar{a}_s, c_s) \, , 
\eeqa
and $a_s := ( \lambda_{l_{a,s}}, \eta_a)$, with similar definitions  for $b_s$ and $c_s$. 
From the above analysis, shifting $1$ and $2$ is one possible valid BCFW shift%
\footnote{In fact shifting any $i$ and $i+1$ would also work when $i$ is odd.}, as indicated in Figure \ref{fig2}. We can then apply the BCFW tree-level recursion relation,
which only affects the tree-amplitude in the top corner of the triple-cut diagram of Figure 
\ref{fig2}, and obtain
\beqa
\label{1234}
 \mathcal{C}_{n; 1, \ldots, m; i } &:=& 
\mathcal{C}( \bar{1}, 2, \ldots, m; \overline{m+1}, \ldots, i; \overline{i+1}, \ldots, n) 
\nonumber \\ \cr
 &=& 
{1\over 2} \sum_{s=1}^2\int d^3 \eta_a d^3 \eta_b d^3 \eta_c \, d^3 \eta_{\hat{P}} 
\nonumber \\ \cr
&&\sum_{k}  {H(z_1, z_2) \over p^2_f } 
\Big[
\mathcal{M}^{(0)}(\bar{3}, \ldots, k, -\bar{\hat{P}}, \hat{2})  
 \mathcal{M}^{(0)}( \overline{k+1}, \ldots, -a_s, -\bar{b}_s, \bar{\hat{1}}, \hat{P} ) \Big]_{z=z_1} 
 \nonumber \\ \cr
&& \times \   
\mathcal{M}^{(0)}(\overline{m+1}, \ldots, i, -\bar{c}_s, b_s ) 
\mathcal{M}^{(0)}(\overline{i+1}, \ldots, n, \bar{a}_s, c_s )
\nonumber \\ \cr
&+ &(z_1 \leftrightarrow z_2 )
\ , 
\eeqa
where $z_{1,2}$ denote the position of the poles in the BCFW recursion relation for the tree amplitude $\mathcal{M}^{(0)}(\bar{1}, \ldots, m, -\bar{b}_s, -a_s) 
$.  These are obtained from \eqref{explsol} with $K_1 = p_{k+1} + \cdots + p_n$, and $K_2 = p_3 + \cdots + p_k$, since $-(p_a+p_b)=p_{m+1}+\cdots+p_n$. 
Note that $z_{1,2}$ and, hence, $\hat{P}$ are independent of the cut loop momenta. 
We should stress that this point is crucial since it implies that the BCFW shifts do not affect
the cut momenta of the triple cut. 
Therefore, we can now rewrite \eqref{1234} as a recursion relation for supercoefficients directly:
\beqa
\label{recsupercoeff}
\mathcal{C}_{n; 1, \ldots, m; i } 
 &=& 
 \sum_{k } (-)^{{m+k \over 2}+1} \int d^3 \eta_{\hat{P}} {H(z_1, z_2) \over p^2_f } 
 \Big[ \mathcal{M}^{(0)}(\bar{3}, \ldots, k, -\bar{\hat{P}}, \hat{2})   \nonumber \\ \cr
 &&\times \   
\mathcal{C}(\bar{\hat{1}}, \hat{P}, \overline{k+1}, \ldots, m ; \overline{m+1}, \ldots, i; \overline{i+1}, \ldots, n)\Big]_{z=z_1} 
\nonumber \\ \cr
& +& 
 (z_1 \leftrightarrow z_2 ) \, ,
\eeqa
where the extra sign factor $(-)^{{m+k \over 2}+1}$ arises from the behaviour of the amplitude under cyclic shifts of its arguments, 
\beq
\mathcal{M}^{(0)} ( \overline{k+1}, \ldots, -a, -\bar{b}, \bar{\hat{1}}, \hat{P} )
= 
(-)^{{m+k \over 2}+1} 
\mathcal{M}^{(0)} (\bar{\hat{1}}, \hat{P}, \overline{k+1}, \ldots, -a, -\bar{b} ) \, .
\eeq 
The reason for this is that we have defined the supercoefficients in such a way that the two cut legs of every tree-level amplitude appear as the last two arguments,  as in \eqref{general}. 
Note that this is  necessary in order to  fix any sign ambiguities.

Eq.~\eqref{recsupercoeff} is the main result of this section. Since in the following sections we will make use  of shifts applied to legs $m-1$ and $m$, we present here the corresponding recursion relation as well, 
\bea
 \mathcal{C}_{n; 1, \ldots, m; i }&=& \sum_{k } (-)^{{k m \over 4} + 1} \int d^3 \eta_{\hat{P}}\, {H(z_1, z_2) \over p^2_f }
\Big[ 
 \mathcal{M}^{(0)}(\bar{k}, \ldots,\overline {\widehat{m\!-\!1} }, \hat{P})  \nonumber \\ \cr
 & \times &  \mathcal{C}(\bar{1}, \ldots, k\!-\!1, -\bar{\hat{P}}, \hat{m} ; \overline{m\!+\!1}, \ldots, i; \overline{i+1}, \ldots, n)\Big]_{z_1} 
 + 
 (z_1 \leftrightarrow z_2 ) \, .
\eea
A key property of the BCFW recursion is that it relates all higher-point amplitudes to the smallest amplitudes in a given theory, which in the case of ABJM can be packaged neatly into the four-point superamplitude.
Therefore we can recursively reduce any corner (of a triple cut) with a higher-point amplitude to a four-point amplitude.
Furthermore, it was shown  recently by the present authors that every triple cut involving  at least one four-point (super)amplitude itself is in one-to-one correspondence with 
a tree-level recursive diagram \cite{btw}. 
Combining these results, we infer that all triple cuts, and hence all one-loop triangle coefficients in ABJM, are effectively related by tree-level recursion relations.
This close relationship between  tree-level recursion  relations and one-loop amplitudes in ABJM   will be explored and made more precise in the following sections. 
We also note note that  this connection makes the Yangian invariance \cite{Gang:2010gy,Bargheer:2010hn} of one-loop amplitudes in ABJM manifest up to the anomalies discussed in \cite{Bargheer:2012cp}.

\section{All one-loop amplitudes}
In this section we illustrate with concrete examples how supercoefficients of one-loop amplitudes in ABJM can be calculated efficiently using the recursion relation formulated in the previous section. In particular all supercoefficients can be related to special ones 
where one of the tree amplitudes in the  triple-cut diagram is a four-point amplitude. 
Since such coefficients can be calculated from an associated tree-level recursive diagram \cite{btw}, this implies that the calculation of all one-loop supercoefficients can be reduced to that of tree recursive diagrams.   The supercoefficients with a four-point amplitude in a corner are therefore the seeds for generic supercoefficients.

\subsection{Six-point amplitude at one corner}

\begin{figure}[h]
\scalebox{1}{
\centerline{\includegraphics[height=6cm]{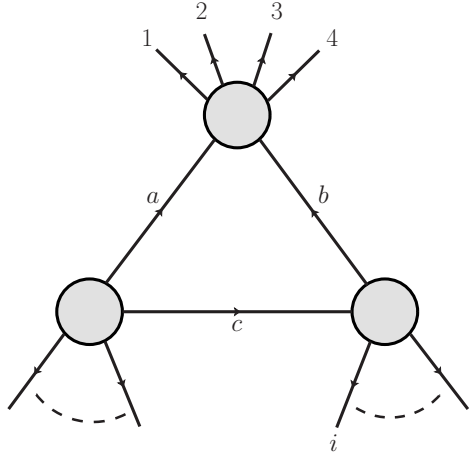} }
}
\caption{\it
A one-loop cut diagram with a six-point tree-level amplitude at one corner. }
 \label{sixptloop}
 \end{figure}

We start with the simplest supercoefficient, when one corner of the associated triple-cut diagram is a six-point tree-level amplitude, as shown in  Figure \ref{sixptloop}. In this case the supercoefficient has the form
\beqa
&&\mathcal{C}_{n; 1234; i } \ =\ {1\over 2} \sum_{s=1}^2  \int\!\!d^3 \eta_a d^3 \eta_b d^3 \eta_c 
\nonumber \\ \cr 
&&\cM^{(0)} (\bar{1},2, \bar{3},4, -\bar{b}_s, -a_s)  
\cM^{(0)}(\bar{5}, \ldots, i, -\bar{c}_s, b_s)  
\cM^{(0)}(\overline{i\!+\!1}, \ldots , n, \bar{a}_s, c_s ) \, . 
\eeqa 
As discussed in Section 3, this supercoefficient can be expressed with a recursion relation with shifts applied to  legs $1$ and $2$  as
\bea \label{6ptloop}
 \mathcal{C}_{n; 1234; i } \ = \  
- \Big[ \cM^{(0)}(\bar{3}, 4, -\bar{\hat{P}}, \hat{2} ) \, \circ \, 
\mathcal{C}_{n-2; {\hat{1}} \, \hat{P}; i } \Big]_{z=z_1}  \, + \, ( z_1\leftrightarrow z_2)
\ ,
\eea
where we have introduced the compact notation  
\beq
\Big[ \cA\circ\cB  \Big]_{z=z_1} \rightarrow  \ \int\!d^3 \eta_{\hat{P}}\, {H(z_1, z_2)\over p_f^2} \Big[ \cA \, \cB \Big]_{z=z_1} 
\ . 
\eeq
%
%
The supercoefficient $\mathcal{C}_{n-2; \hat{1} \, \hat{P}; i }$ is obtained from a triple-cut diagram where one of the participating tree amplitudes is a four-point amplitude. As we reviewed earlier, these particular supercoefficients can be  obtained from certain  associated BCFW tree-level diagrams, denoted as $Y$ in the previous sections, and  the precise correspondence   is given in \eqref{supxm}. 
Using this result, we can rewrite $ \mathcal{C}_{n; 1234; i }$ as    
\bea
\label{Ynew}
 \mathcal{C}_{n; 1234; i }  
& =&
 -{1\over 4} \lan \hat{1} \hat{P} \ran \lan \xi_{5i} \mu_{5i} \ran \lan \xi_{i+1 \, n} \mu_{i+1 \, n} \ran \  
 \Big[ \cM(\bar{3}, 4, -\bar{\hat{P}}, \hat{2} ) \circ
 \big( Y^{(1)}_{n-2;\hat{1} \hat{P}; i } - Y^{(2)}_{n-2;\hat{1} \hat{P}; i } \big )   \Big]_{z_1}
  \nonumber \\ \cr
  & + & (z_1 \leftrightarrow z_2)
  \, ,
\eea
where a specific choice of signs has been made to  rewrite massive momenta in terms of spinors (see \eqref{K1K2}), 
\beq
\sqrt{ P^2_{5i} P^2_{i+1 \, n}} \ = \ - (1/4) \lan \xi_{5i} \mu_{5i} \ran \lan \xi_{i+1 \, n} \mu_{i+1 \, n} \ran
\,  . 
\eeq 
In \eqref{Ynew} we have also  introduced a more informative notation for the $Y$-functions, in that we now specify the total number of legs
($n-2$ in the present case). 
This  is because  when  repeatedly applying recursion relations, we have to deal with several  $Y$-functions 
with different number of  legs at the same time. For instance, the $Y$-function appearing in  \eqref{Ynew} is 
\bea \label{1ppole}
Y^{( \alpha )}_{n-2;\hat{1} \hat{P}; i } 
:=
\Big[ \cM^{(0)}(\bar{5}, \ldots, i, -\bar{\hat{P}}_{1P}, \hat{\hat{P}})  \circ 
\cM^{(0)}(\overline{i+1}, \ldots, n, \hat{\hat{1}}, \hat{P}_{1P}) \Big]_{z^{\hat{1} \hat{P}}_{\alpha }} \, ,
\eea 
with $\alpha = 1, 2$, and where $\hat{P}_{1P}$ denotes the propagator of the BCFW diagram with   legs  $\hat{1}$,  $\hat{P}$ being  shifted.
We have also added a superscript to the pole solutions $z_\alpha$ -- for instance  $z^{\hat{1} \hat{P}}_{\alpha }$ refers to a recursive diagram with  shifted legs  $\hat{1}$ and $\hat{P}$. 

We also find it convenient to introduce a more general notation  
 $Y^{( \alpha )}_{n; i \, i+1; k }$ 
in order to capture  the result of an iterated recursion. To this end we define the functions
\bea \label{Yfunction}
Y^{(\alpha_1 ; \alpha_2 ; \ldots; \alpha_m)}_{n; i_{1_1} i_{1_2}; i_{2_1} i_{2_2}; \ldots ; i_{m_1} i_{m_2}; k_1; k_2; \ldots ; k_m } \, ,
\eea
where $n$ is again the total number of legs, and $i_{j_1} i_{j_2}$ indicate which adjacent pair of legs are shifted when we apply the  $j^{\rm th}$  recursion relation. Finally,  for each recursive diagram with shifts $i_{j_1} i_{j_2}$, we specify the corresponding channel by adding an index $k_j$, similarly to the index  $k$ in  $Y^{( \alpha )}_{n; i \, i+1; k }$. 
 The functions $Y$ so defined appear in  iterated  BCFW recursions of tree amplitudes and are the natural building blocks of one-loop supercoefficients.

Making use of this more compact notation, we find%
\footnote{In order to  perform this further shift on $\lambda_{\hat{P}}$ explicitly, one may use its explicit expression given in \eqref{onshellP}.} 
\bea 
 &&Y^{(\alpha ; \beta)}_{n; 12; \hat{1} \hat{P}; 4, i}  
 \ := \
\Big[ \cM^{(0)}(\bar{3}, 4, -\bar{\hat{P}}, \hat{2} ) 
 \circ 
 Y^{(\beta)}_{n-2;\hat{1} \hat{P}; i }\Big]_{z^{12}_{ \alpha }} \\ \nonumber \cr
 &&= 
\Big[  \cM^{(0)}(\bar{3}, 4, -\bar{\hat{P}}, \hat{2} )\circ
\big[ 
 \cM^{(0)}(\bar{5}, \ldots, i, -\bar{\hat{P}}_{1P}, \hat{\hat{P}}) \circ 
 \cM^{(0)}(\overline{i+1}, \ldots, n, \hat{\hat{1}}, \hat{P}_{1P})  \big]_{z^{\hat{1} \hat{P}}_{\beta }}  
\Big]_{z^{12}_{\alpha }}
  \, .
\eea
Thus we can rewrite the one-loop supercoefficient  as
\bea
\label{111}
\mathcal{C}_{n; 1234; i } 
 = 
 {1 \over 4} \lan \xi_{5i} \mu_{5i} \ran \lan \xi_{i+1 \, n} \mu_{i+1 \, n} \ran  
 \sum^{2}_{ \alpha, \beta = 1 } (-)^{\beta } \,
\lan \hat{1} \, \hat{P} \ran  Y^{(\alpha; \, \beta)}_{n;12; \hat{1} \hat{P};4; i } \, .
\eea
It is easy to evaluate $\lan \hat{1} \, \hat{P} \ran$ explicitly   using  the solution for 
$\lambda_{\hat{P}}$ in  \eqref{onshellP},  with the result 
\bea
\left. \lan \hat{1} \, \hat{P} \ran \right |_{z^{12}_1} = {i \over 2} \lan \xi_{14} \mu_{14} \ran \, ,  \qquad 
\left.  \lan \hat{1} \, \hat{P} \ran  \right|_{z^{12}_2} = -{i \over 2} \lan \xi_{14} \mu_{14} \ran \, .
\eea 
Using this,  \eqref{111} becomes
\bea
 \mathcal{C}_{n; 1234; i } 
  = 
  {i \over 8} \lan \xi_{14} \mu_{14} \ran \lan \xi_{5i} \mu_{5i} \ran \lan \xi_{i+1 \, n} \mu_{i+1 \, n} \ran  
 \sum^{2}_{ \alpha, \beta = 1 } (-)^{\alpha + \beta + 1 } \,
 Y^{(\alpha; \, \beta)}_{n;12; \hat{1} \hat{P};4; i }   \, .
\eea
Multiplying $ \mathcal{C}_{n; 1234; i }$ with  the corresponding three-mass triangle we  obtain the contribution of this diagram to the one-loop amplitude, 
\bea
 \mathcal{C}_{n; 1234; i } \cI_{1,4; 5,i; i+1,n}  
\  = \ 
 \mathcal{S}_{1,4; 5,i; i+1,n} \, 
 \sum^{2}_{ \alpha, \beta = 1 } (-)^{\alpha + \beta + 1 } 
 Y^{(\alpha; \, \beta)}_{n;12; \hat{1} \hat{P};4; i }   \, ,
\eea
where the pre-factor $\mathcal{S}_{1,4; 5,i; i+1,n}$ is%
\footnote{For the sake of clarity, we remind the reader that  $Y^{(\alpha; \, \beta)}_{n;12; \hat{1} \hat{P};4; i }$ is a tree recursive diagram which would appear in the tree-level amplitude $\cM^{(0)}_{n}(\bar{1},2,  \ldots, n )$, namely 
$
\sum^{2}_{\alpha, \, \beta =1 } Y^{(\alpha; \beta)}_{n;12; \hat{1} \hat{P}; 4; i } \in 
\cM^{(0)}_{n}(\bar{1},2,  \ldots, n )$.
} 
\bea
\mathcal{S}_{1,4; 5,i; i+1,n} 
\, = \, 
{ \pi^3  \over 8 } {\lan \xi_{14} \mu_{14} \ran \xi_{5i} \mu_{5i} \ran \lan \xi_{i+1 \, n} \mu_{i+1 \, n} \ran   \over  \sqrt{ -(P^2_{1,4} + i \varepsilon) } \sqrt{ -( P^2_{5,i} + i \varepsilon) }  
\sqrt{ -( P^2_{i+1,n} + i \varepsilon ) } } \, .
\eea  


\subsection{Eight-point amplitude at one corner}

Before discussing  the case of a generic  one-loop supercoefficient, we consider a sightly more sophisticated example in detail, specifically the situation  where a  tree-level amplitude at one corner  of the triple-cut diagram is an eight-point amplitude. Such a supercoefficient can be written as 
\beqa
&& \cC_{n; 1, \ldots, 6; i} \   = \   {1\over 2} \sum_{s=1}^2 
\int\!\!d^3 \eta_a  d^3 \eta_b d^3 \eta_c
\nonumber   \\ \cr
&&\cM^{(0)}(\bar{1}, \ldots, 6, -\bar{b}_s, -a_s) 
\cM^{(0)}(\bar{7}, \ldots, i, -\bar{c}_s, b_s) 
\cM^{(0)}(\overline{i+1}, \ldots, n, \bar{a}_s, c_s) 
  \, .
\eeqa
We can derive this from a recursion relation with legs  $1$ and $2$ shifted: 
\bea
\label{cc4}
\mathcal{C}_{n; 1, \ldots, 6; i} \ &=& \ 
- \Big[ \cM^{(0)}( \bar{3}, \ldots, 6, -\bar{\hat{P}}, \hat{2}) \, 
\circ \mathcal{C}_{n-4; \hat{1} \hat{P}; i} \Big]_{z^{12}_1}  
+ 
\Big[ \cM^{(0)}(\bar{3}, 4, -\bar{\hat{P}}, \hat{2} ) \, 
\circ \mathcal{C}_{n-2;\hat{1} \hat{P} 5 6; i} \Big]_{z^{12}_1} \nonumber \\ \cr
& + & (z^{12}_1 \leftrightarrow z^{12}_2 )  \, . 
\eea
The first term on the right-hand side of \eqref{cc4} needs no further reduction as it already contains a supercoefficient which can be derived using the correspondence with tree-level recursive diagrams of  \cite{btw}.  On the other hand, for the second term we can apply once more the  supercoefficient recursion relation,  shifting for instance the legs $5$ and $6$ of $\cC_{n-2; \hat{1} \hat{P} 56; i}$. 
To distinguish the new shifts from those of the  first recursion relation, 
we denote them as  $\hat{5^\prime}$, $\hat{6^\prime}$,  and the  propagator in the corresponding recursive diagram as  $\hat{P^\prime}$. 
After applying this second   recursion, $\mathcal{C}_{n; 1, \ldots, 6; i}$  can be written as  
\bea 
\label{eight}
\mathcal{C}_{n; 1, \ldots, 6; i} 
&=& 
-\Big[ \cM^{(0)}( \bar{3}, \ldots, 6, -\bar{\hat{P}}, \hat{2}) \, \circ
\mathcal{C}_{n-4; \hat{1} \hat{P}; i}  \Big]_{z^{12}_1}
\nonumber  \\ \cr 
&&+  \, 
\bigg[
\Big[ \cM^{(0)}(\bar{3}, 4, -\bar{\hat{P}}, \hat{2} ) 
\circ \big[ \cM^{(0)}(\bar{\hat{1}}, \hat{P}, \bar{\hat{5^\prime}}, \hat{P^\prime}  ) 
\circ (-\mathcal{C}_{n-4; -\hat{P^\prime} \hat{6^\prime} ; i}) \big]_{z^{56}_1} \, \Big]_{z^{12}_1}  \nonumber \\ \cr
& &+ \, (z^{56}_1  \leftrightarrow  z^{56}_2)
\bigg] + ( z^{12}_1 \leftrightarrow  z^{12}_2 )   \nonumber \\ \cr
&=& -\Big[ \cM^{(0)}( \bar{3}, \ldots, 6, -\bar{\hat{P}}, \hat{2}) 
\circ \mathcal{C}_{n-4; \hat{1} \hat{P}; i}  \Big]_{z^{12}_1}
\nonumber \\ \cr
&& +\,  
\Big[ \cM^{(0)}(\bar{1}, 2, \bar{3}, 4, \bar{\hat{5^\prime}}, \hat{P^\prime}  ) 
\circ \mathcal{C}_{n-4; -\hat{P^\prime} \hat{6^\prime} ; i}  \Big]_{z^{56}_1}  
\nonumber \\ \cr
& &+ \,  (z^{56}_1  \leftrightarrow  z^{56}_2) + ( z^{12}_1 \leftrightarrow  z^{12}_2 )  \, ,
\eea
where in the last step we used 
\bea
\Big[ \cM^{(0)}(\bar{3}, 4, -\bar{\hat{P}}, \hat{2} ) 
\circ \cM^{(0)}(\bar{\hat{1}}, \hat{P}, \bar{\hat{5'}}, \hat{P'}  )  \Big]_{z^{12}_1} 
+ (z^{12}_1 \leftrightarrow  z^{12}_2)
= -\cM^{(0)}(\bar{1}, 2, \bar{3}, 4, \bar{\hat{5^\prime}}, \hat{P^\prime}  )  .
\eea
Now we have reached the extremal case when all the supercoefficients have at least one corner with a four-point amplitude, so we are ready to apply the connection between these special  triple-cuts and tree-level recursion relations, and find
\bea
\label{098}
\mathcal{C}_{n; 1, \ldots, 6; i} &=& 
{1 \over 4}\lan \xi_{7i} \mu_{7i} \ran  \lan \xi_{i+1 n} \mu_{i+1 n} \ran \nonumber \\ \cr
&&\times \, 
\Big( - \lan \hat{1} \, \hat{P} \ran  
\big[ \cM^{(0)}( \bar{3}, \ldots, 6, -\bar{\hat{P}}, \hat{2}) \circ 
(Y^{(1)}_{n-4; \hat{1} \hat{P}; i} 
-
 Y^{(2)}_{n-4; \hat{1} \hat{P}; i}) \big]_{z^{12}_1} \nonumber \\ \cr 
&&+\,   
i\lan \hat{P} \, \hat{6} \ran 
\big[ \cM^{(0)}(\bar{1}, 2, \bar{3}, 4, \bar{\hat{5}}, \hat{P}  ) \circ
 ( Y^{(1)}_{n-4; -\hat{P} \hat{6}; i} 
 -
  Y^{(2)}_{n-4; -\hat{P} \hat{6}; i} ) \big]_{z^{56}_1} \Big) \nonumber \\ \cr
  & &+\,  (z^{56}_1  \leftrightarrow  z^{56}_2) + ( z^{12}_1 \leftrightarrow  z^{12}_2 ) 
  \, ,
\eea
where in the second line  the prime on the shifted legs $5$ and $6$ has been removed, since in this expression we only perform BCFW shifts on legs $5$ and $6$, but not legs $1$ and $2$. Various terms in \eqref{098} can be combined  into the $Y$-functions defined in  \eqref{Yfunction}. After some simple manipulations we obtain
\bea
\mathcal{C}_{n; 1, \ldots, 6; i}
\, =  \,  
{i \over 8}\lan \xi_{16} \mu_{16} \ran \lan \xi_{7i} \mu_{7i} \ran  \lan \xi_{i+1 n} \mu_{i+1 n} \ran \sum^2_{\alpha, \, \beta=1} (-)^{\alpha + \beta + 1} ( Y^{(\alpha; \beta)}_{n;12; \hat{1} \hat{P};6; i} + Y^{(\alpha; \beta)}_{n;56; -\hat{P} \hat{6};n; i}) \, , 
\eea
where we have used the fact that  $\lan \hat{1}  \hat{P} \ran$ and $i \lan \hat{P} \hat{6}  \ran$ can be written in terms of 
$\lan \xi_{16} \mu_{16} \ran$. 
Multiplying $\mathcal{C}_{n; 1, \ldots, 6; i}$  with the corresponding three-mass triangle integral, we obtain the one-loop contribution of this diagram, 
\bea \label{8ptloop}
\mathcal{C}_{n; 1, \ldots, 6; i} I_{1, 6; 7, i; i+1, n}\ =\ 
\mathcal{S}_{1, 6; 7, i; i+1, n} \sum^2_{\alpha, \, \beta=1} (-)^{\alpha + \beta + 1} ( Y^{(\alpha; \beta)}_{n;12; \hat{1} \hat{p};6; i} + Y^{(\alpha; \beta)}_{n;56; -\hat{P} \hat{6};n; i})   \, ,
\eea
where the pre-factor  $\mathcal{S}_{1, 6; 7, i; i+1, n}$ is
\bea
\mathcal{S}_{1, 6; 7, i; i+1, n}
 \, = \, 
{ \pi^3 \over 8} { \lan \xi_{16} \mu_{16} \ran \lan \xi_{7i} \mu_{7i} \ran  \lan \xi_{i+1 n} \mu_{i+1 n} \ran \over \sqrt{- (P^2_{1, 6}+i\varepsilon)} \sqrt{- (P^2_{7, i}+i\varepsilon)} \sqrt{- (P^2_{i+1, n}+i\varepsilon)}   }\, .
\eea

\subsection{The general one-loop supercoefficients}

It is not difficult to generalise the results of the previous subsections to arbitrary supercoefficients such as 
\eqref{general}.  Without loss of generality we  focus on the tree-level amplitude $\cM^{(0)}(\bar{1}, \ldots, m, -\bar{b}, -a)$. 
The idea is to repeatedly apply the recursion relation in order to reduce it to a four-point amplitude, and then apply the connection between anomalous triple cuts and tree-level recursion diagrams. 

As in  the special cases of the previous subsections, we start by  shifting legs $1$ and $2$, followed  by shifts of legs $m\!-\!1$ and $m$. 
After that, we shift the  two left-most legs of that corner, namely the shifted leg $1$ and the corresponding shifted propagator. Generally we denote them as $\hat{1}_q$ and $\hat{P}_q$, {\it i.e.} we call them as $\hat{1}_1$ and $\hat{P}_1$ when they are shifted in the first iteration, and $\hat{1}_2$ and $\hat{P}_2$ in the second iteration, and so on.  This process terminates whenever we reach the extremal case, namely when an $(m+2)$-point amplitude at the corner reduces to a four-point amplitude. 
The result of this procedure  is 
\bea  \label{generaloneloop}
\mathcal{C}_{n; 1, 2, \ldots, m; i} &=& 
{i \over 8} \lan \xi_{1m} \mu_{1m} \ran \lan \xi_{m+1 i} \mu_{m+1 i} \ran  \lan \xi_{i+1 n} \mu_{i+1 n} \ran 
 \\ \cr
&&\times\  
\! \!
\big( \sum^2_{ \alpha, \beta = 1 } (-)^{\alpha + \beta + 1} \big(  Y^{(\alpha; \beta )}_{n; 1 \, 2; 1_1 \, P_1; m; i}
+ Y^{(\alpha; \beta )}_{n; m-1 \, m; -P_1 \, m_1; n; i} 
 \big) \nonumber \\ \cr
 &&+ \
 \! \!  \sum_{j, \alpha, k} (-)^{{k_1 + \ldots + k_{j-1} \over 2 }+{m (j+1) \over 2} + \alpha_{0} + \alpha_{j+1}} Y^{(\alpha_0; \alpha_1; \ldots ; \alpha_{j+1} ) }_{n;12; 1_1 P_1; \ldots ; 1_{j+1} P_{j+1}; k_0; k_1 ; \ldots ; k_{j-1}; m ; i}   \big) ,
\nonumber
\eea
where the summation in the last term is over $j$, the $\alpha$'s, which can be $1$ and $2$, and all the $k$'s, which must be even with $k_p < k_q$ if $p < q$. In order to arrive at the expression  in the last line of  \eqref{generaloneloop} we used the following identity
\bea
&& \Big[ 
\cM^{(0)}(\overline{k_j\!+\!1}, \ldots, k_f, -\bar{\hat{P}}_{m-1, m}, \hat{m} ) \circ
\cM^{(0)}(\overline{k_f\!+\!1}, \ldots, \widehat{m\!-\!1}, \hat{P}_{m-1, m} )
\Big]_{z_1} + (z_1  \leftrightarrow z_2 ) \nonumber \\ \cr 
&=& 
\cM^{(0)}(\overline{m\!-\!1 }, m, \hat{P}_{1_j P_j}, \hat{P}_j, \overline{k_j\!+\!1}, \ldots, k_f, \overline{k_f\!+\!1}, \ldots, m-2  ) \nonumber \\ \cr
&=&
\cM^{(0)}(\overline{k_j\!+\!1}, \ldots, k_f, \overline{k_f\!+\!1}, \ldots, m-2, \overline{m\!-\!1 }, m, \hat{P}_{1_j P_j}, \hat{P}_j  ) \, .
\eea
To obtain the contribution to the one-loop amplitude we simply multiply the supercoefficient with the corresponding triangle integral, with the result
\beqa
\label{generaloneloop2}
&& \mathcal{C}_{n; 1, 2, \ldots, m; i} \cI_{1,m; m+1, i; j+1, n} \ = \  \mathcal{S}_{1,m; m+1, i; j+1, n}
\nonumber \\ \cr
&&\times \
\Big[ 
\sum^2_{ \alpha, \beta = 1 } (-)^{\alpha + \beta + 1 } \big(  Y^{(\alpha; \beta )}_{n; 1 \, 2; 1_1 \, P_1; m; i}
+ Y^{(\alpha; \beta )}_{n; m-1 \, m; -P_1 \, m_1; n; i} 
 \big) 
\nonumber \\ \cr
 &&+\ 
 \sum_{j, \alpha, k} (-)^{{k_1 + \ldots + k_{j-1} \over 2 }+{m (j+1) \over 2 } + \alpha_{0} + \alpha_{j+1}} Y^{(\alpha_0; \alpha_1; \ldots ; \alpha_{j+1} ) }_{n;12; 1_1 P_1; \ldots ; 1_{j+1} P_{j+1}; k_0; k_1 ; \ldots ; k_{j-1}; m ; i}   \Big] \, , 
\eea
where
\bea
\mathcal{S}_{1,m; m+1, i; j+1, n} =  {\pi^3 \over 8} { \lan \xi_{1m} \mu_{1m} \ran \lan \xi_{m+1 i} \mu_{m+1 i} \ran  \lan \xi_{i+1 n} \mu_{i+1 n} \ran   \over \sqrt{ -(P^2_{1, m}+i\varepsilon)} \sqrt{ -(P^2_{m+1, i}+i\varepsilon)} \sqrt{ -(P^2_{i+1, n}+i\varepsilon)} } \, .
\eea
We wish to    emphasise that by definition  each sum of $Y$-functions (without the minus signs) in \eqref{generaloneloop2} 
is equal to terms which would appear in iterated BCFW recursion relations for tree amplitudes. 
In this sense,  all one-loop amplitudes can  be written as sums of tree-level recursive diagrams with possible minus signs. We also note that  
 each term in this sum is dual conformal invariant. 



\section*{Acknowledgements}

We would like to thank Niklas Beisert, Paul Heslop, Valya Khoze and Bill Spence for  stimulating  discussions. 
This work was supported by the STFC Grant ST/J000469/1,  
``String theory, gauge theory \& duality". 
AB thanks the Physics Departments at the Weizmann Institute of Science and
Tel Aviv University for their hospitality. GT acknowledges the warm
hospitality and support from the Institute for Particle Physics
Phenomenology, Durham University,  through an IPPP Associateship. We would also like to thank the ECT*, Trento, for hospitality and partial support.


\appendix

\section{On-shell solution}

\begin{figure}[h]
\scalebox{1}{
\centerline{\includegraphics[height=4.5cm]{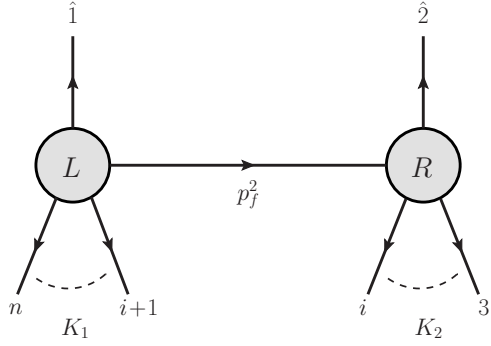} }
}
\caption{\it
A BCFW recursive diagram of a tree-level amplitude. Here $K_1 = p_{i+1} + \ldots + p_n$ and 
$K_2 = p_{3} + \ldots + p_i$.}
 \label{fig:bcfw}
 \end{figure}

We present here a solution for the spinor  $\lambda_{\hat{P}}$ that appeared in the the BCFW recursion relations. In general, 
any two-dimensional vector can be expanded in terms of two independent vectors. In our case it is convenient to expand $\lambda_{\hat{P}}$ in terms of $\lambda_{\hat{1}}$ and $\lambda_{\hat{2}}$, as
\bea
\label{sop}
\lambda_{\hat{P}} = {1 \over  \lan 1 \, 2\ran } \Big( \lan \hat{P}  \, \hat{2} \ran   \lambda_{\hat{1}} +  \lan \hat{1}  \, \hat{P} \ran \lambda_{\hat{2}}  \Big) \, ,
\eea
where we have used that  $\lan \hat{1}  \, \hat{2}\ran = \lan 1 \, 2\ran$.  \eqref{sop} can be  further simplified by applying  momentum conservation, see Figure \ref{fig:bcfw}, 
\bea 
\lan \hat{P}  \, \hat{1} \rangle^2 &=& (\hat{P}  + \hat{1})^2 = K^2_1 
= -{1 \over 4} \lan \xi_{i+1 \, n} \, \mu_{i+1 \, n} \ran^2 \, , 
\\ \nonumber \cr  
\lan \hat{P}  \, \hat{2} \rangle^2 &=& -(\hat{P}  - \hat{2})^2 = -K^2_2 
= {1 \over 4} \lan \xi_{3 i} \, \mu_{3 i} \ran^2 \, ,
\eea
where we have rewritten massive momenta in terms of spinors. Thus we arrive at the result 
\bea \label{onshellP}
\lambda_{\hat{P}}\  =\  \pm  \, {1 \over  2\lan 1 \, 2\ran } 
\Big(  \lan \xi_{3 i} \, \mu_{3 i} \ran \,   \lambda_{\hat{1}} \, 
\pm\,  i \lan \xi_{i+1 \, n} \, \mu_{i+1 \, n} \ran \, \lambda_{\hat{2}}  \Big)  \, ,
\eea
where the four possible choices of signs  correspond  to the four possible BCFW on-shell solutions,  
\beq
z^2_1 \ =  \ {  \lan \xi_{i+1 \, n} \mu_{3 i} \ran \lan \mu_{i+1 \, n} \xi_{3 i} \ran  \over \lan \l_1 + i \l_2 | \, K_1\, | \l_1 + i \l_2 \ran  }
 \,  , \qquad 
z^2_2 \ =  \ { \lan \xi_{i+1 \, n} \xi_{3 i} \ran  \lan \mu_{i+1 \, n} \mu_{3 i} \ran \over \lan \l_1 + i \l_2 | \, K_1\, | \l_1 + i \l_2 \ran }
\ .
\eeq


\end{document}